\documentclass[aps,prl,twocolumn,groupedaddress,showpacs,nofootinbib]{revtex4}
\usepackage{graphicx}
\usepackage{epstopdf}
\usepackage{latexsym}
\usepackage{amsfonts}
\usepackage{amssymb}
\usepackage{amsmath}
\usepackage{slashed}
\usepackage{feynmp}

\begin{document}

\title{High-Energy Neutrino Signatures of Dark Matter Decaying into Leptons}
\author{Matthew R. Buckley$^1$,  Katherine Freese$^2$, Dan Hooper$^{3,4}$, 
Douglas Spolyar$^{3,5}$, and Hitoshi Murayama$^{6,7,8}$}
\affiliation{$^1$Department of Physics, California Institute of Technology, Pasadena, CA 91125, USA}
\affiliation{$^2$Michigan Center for Theoretical Physics,
 Physics Dept.,  University of Michigan, Ann Arbor, MI 48109, USA}
\affiliation{$^3$Center for Particle Astrophysics, Fermi National Accelerator Laboratory, Batavia, IL 60510 USA}
\affiliation{$^4$Astronomy and Astrophysics Department, University of Chicago, Chicago, IL 60637 USA}
\affiliation{$^5$University of California, Santa Cruz, Physics Department,
Santa Cruz, CA 95064 USA}
\affiliation{$^6$Department of Physics, University of California, Berkeley, CA 94720, USA}     
\affiliation{$^7$Theoretical Physics Group, LBNL, Berkeley, CA 94720, USA}
\affiliation{$^8$IPMU, University of Tokyo, 5-1-5 Kashiwa-no-ha, Kashiwa,
                Japan 277-8568}
                \date{\today}

\begin{abstract}
Decaying dark matter has previously been proposed as a possible
explanation for the excess high energy cosmic ray electrons and positrons seen by PAMELA and
the Fermi Gamma-Ray Space Telescope (FGST). To accommodate these signals however, the decays must be predominantly leptonic, to muons or taus, and therefore
produce neutrinos, potentially detectable with the IceCube neutrino observatory.  We find
that, with five years of data, IceCube (supplemented by DeepCore) will be able to significantly constrain the relevant parameter space of decaying dark matter, and may even be capable of discovering dark matter decaying in the halo of the Milky Way. 
\end{abstract}

\pacs{95.35.+d;95.30.Cq,98.52.Wz,95.55.Ka;FERMILAB-PUB-09-344-A;CALT-68-2744}

\maketitle

A number of recent measurements of cosmic ray electrons and positrons have been interpreted as possible evidence for dark matter~\cite{Adriani:2008zr,Chang:2008zzr,Torii:2008xu,Collaboration:2009zk,Barwick:1997ig,ams}. In particular, it has been suggested that the features of the $e^+e^-$ spectrum and positron fraction reported by PAMELA \cite{Adriani:2008zr}, ATIC \cite{Chang:2008zzr}, PPB-BETS \cite{Torii:2008xu}, the Fermi Gamma Ray Space Telescope (FGST) \cite{Collaboration:2009zk}, HEAT \cite{Barwick:1997ig}, and AMS-01\cite{ams} may originate from either DM annihilations \cite{annihilation,weiner,Bergstrom:2009fa,leptonic,Sommerfeld1,Sommerfeld2} or decays \cite{decay} taking place in the Galactic Halo. It is also possible that these signals originate from astrophysical sources, rather than from new high energy physics \cite{Profumo:2008ms}. 

Although annihilating dark matter could potentially generate the observed anomalous cosmic ray features, attempts to do so face a number of challenges. Firstly, the spectrum of electrons and positrons predicted to be generated in the annihilations of most dark matter candidates is much too soft to fit the observations of PAMELA and FGST~\cite{weiner,Bergstrom:2009fa}. If WIMPs annihilating throughout the halo of the Milky Way are to produce the spectral shape observed by these experiments they must annihilate mostly to charged leptons. While models have been proposed in which this is the case~\cite{leptonic,Sommerfeld1}, many of the most often studied WIMP candidates (including MSSM neutralinos) are predicted to annihilate dominantly to quarks and/or gauge bosons~\cite{Primack:1988zm}. Furthermore, annihilations to non-leptonic final states tend to produce more cosmic ray anti-protons than are observed~\cite{antiprotons}. Secondly, the dark matter annihilation rate that is required to generate the observed spectrum of cosmic ray electrons and positrons is considerably higher than is predicted for a typical thermal relic distributed smoothly throughout the Galactic halo. To normalize the annihilation rate to the PAMELA and FGST signals, we must require either large inhomogeneities in the dark matter distribution which lead to a considerably enhanced annihilation rate ({\it i.e.}~a ``boost factor''), and/or dark matter particles which possess a considerably larger annihilation cross section than is required of a thermal relic. This latter requires either a non-thermal production mechanism in the early universe, or an enhancement of the annihilation cross section at low velocities, such as through the Sommerfeld effect~\cite{Sommerfeld1,Sommerfeld2} or Breit-Wigner enhancement~\cite{BW}. In this light of these challenges, the observations from PAMELA and FGST are extremely surprising and pose an interesting challenge to the usual WIMP paradigm.

In this paper we consider decaying dark matter as a possible origin of these excess electrons and positrons, as has previously been discussed in Refs.~\cite{Ishiwata:2009vx,Meade:2009iu,Ibarra:2009dr}. 
The only properties of the dark matter relevant to this work are its mass, its lifetime, and its decay channels.  Similar to the discussion in the previous paragraph, any explanation of the PAMELA and FGST data requires preferentially leptonic decay products.  Beyond these few phenomenological properties, the nature of the decaying particles is not relevant to our study. 

In order to confirm a particle physics origin of the PAMELA and FGST data, one would hope to observe other final state particles in addition to electrons and positrons. It has been shown that dark matter with a mass $\sim10^{2-5}$~GeV that annihilates/decays into $\mu^-\mu^+$, $\tau^-\tau^+$, or $\mu^-\mu^+\mu^-\mu^+$ can reproduce the observed cosmic ray features~\cite{Bergstrom:2009fa}. In such a scenario, we also expect the associated production of gamma rays and neutrinos from the decay of the heavy charged leptons. 

In a previous letter \cite{Spolyar:2009kx}, the authors examined the sensitivity of the IceCube detector to neutrinos from dark matter pair-annihilation in the inner Milky Way. As the annihilation rate rises with the dark matter density squared, the inner galaxy is a promising region to observe such signatures. Unfortunately, at the Antarctic location of IceCube, the Galactic center is overhead, causing the signal to be swamped with background from atmospheric muons. However, the DeepCore extension of IceCube (to be completed in 2010) \cite{Klein:2008px,Resconi:2008fe}, the heavily instrumented inner volume of the detector can make use of the remaining IceCube volume to reject the muon background, allowing for neutrino-induced showers to be identified. This reduces the background to those events from atmospheric neutrinos, which is not overwhelming compared to the expected signal. Using this technique, it will be possible to place bounds on the dark matter annihilation rate that are comparable to those required to explain the PAMELA/FGST results.

A natural extension of this idea is to consider dark matter decay. In this letter, we consider dark matter decaying into muon pairs, tau pairs, or four muon final states, each of which can adequately explain the observed electron/positron spectrum, for dark matter lifetimes on the order of $\tau\sim 10^{26-27}$~seconds~\cite{Ishiwata:2009vx,Meade:2009iu,Ibarra:2009dr}.  
In all three of these cases, the subsequent decays of the muons and taus will create copious numbers of high energy neutrinos, potentially detectable at IceCube.

The flux of neutrinos from dark matter decay in the inner Milky Way is given by
\begin{equation}
\frac{d\Phi(\Delta \Omega,E)}{dE} = \frac{1}{4\pi}\frac{\Gamma}{m_\chi} \bar{{\cal J}}(\Delta \Omega)\Delta \Omega\sum_i \frac{dN_i}{dE}.\label{eq:dPhidE}
\end{equation}
Here, $\Gamma = \tau^{-1}$ is the decay width of the dark matter, $m_\chi$ is the dark matter mass, and $dN_i/dE$ is the differential flux of neutrinos of flavor $i$ resulting from the decay. The dark matter distribution integrated over the line-of-sight over a solid angle $\Delta \Omega$ is given by
 \begin{equation}
 {\cal J} = \int_{l.o.s.} \rho_\chi(s) ds \hspace{1em};\hspace{1em}
 \bar{{\cal J}}(\Delta \Omega) = 
 {1 \over \Delta \Omega} 
 \int_{\Delta \Omega} PSF \star {\cal J} d\Omega \label{eq:Jdef}
 \end{equation}
 where PSF is the point spread function of the instrument and $\rho_\chi(s)$ is the dark matter mass density distribution. Note that this differs from the definition of $\bar{J}$ used in Ref.~\cite{Spolyar:2009kx}, in which $\bar{J} \propto \rho^2_\chi$. As we are interested here in decay rather than annihilation, only one power of the density appears.
 
The spectrum of neutrinos from the decays depends on the mass of the dark matter and on the dominant decay channel. For muon-channel decays ($\mu^+\mu^-$ and $\mu^+\mu^-\mu^+\mu^-$), the muons themselves decay to $\nu_\mu e \nu_e$. For decays through taus, there are many final states available, including $\tau \rightarrow \mu \nu_\mu \nu_\tau$, $e \nu_e \nu_\tau$, as well as from the hadronic decays $\tau \rightarrow \pi \nu_\tau$, $K \nu_\tau$, $\pi \pi \nu_\tau$, and $\pi \pi \pi \nu_\tau$~\cite{Jungman:1994jr}. For this work, the spectrum of neutrinos from tau decays as a function of $m_\chi$ was determined numerically using Pythia \cite{Sjostrand:2006za}. It should be noted that for the four muon final state, it is assumed that the two $\mu^+\mu^-$ pairs originate from the decays of back-to-back parent particles. As a result, the rest energy $m_\chi$ is split evenly between the two pairs, rather than distributed among the four particles as per naive phase-space.
 
The primary backgrounds for this observation consist of atmospheric muons and neutrinos. As addressed previously, the IceCube detector itself can be used to veto muons inside of the volume of DeepCore, leaving only neutrino-induced showers to compete with. For the spectrum of atmospheric muon neutrinos, we use the results of Ref.~\cite{Honda:2006qj}, which are in good agreement with the measurements of AMANDA~\cite{Collaboration:2009nf}.\footnote{As the $\nu_\mu$ background
is considerably larger than that from $\nu_e$'s, any discrimination between electromagnetic and hadronic showers could be used to further reduce the backgrounds and improve the statistical reach of IceCube/Deepcore to the signal described in this paper.}

The effective area of the detector for neutrinos can be defined as
\begin{equation}
A(E) \approx \rho_{\rm ice} N_A \sigma_{\nu N}(E)V(E),\label{eq:effarea}
\end{equation}
where $\rho_{\rm ice} =0.9 ~\mbox{g/cm}^3$, $N_A = 6.022 \times 10^{23}~\mbox{g}^{-1}$ (to convert grams to nucleons), $\sigma_{\nu N}(E)$ is the neutrino-nucleon cross-section~\cite{Gandhi:1998ri} and $V(E) \approx 0.04$ km$^3$ is the effective volume of the DeepCore detector for a neutrino-induced shower of energy $E$~\cite{Resconi:2008fe}.

The directional capability of IceCube/DeepCore for a neutrino-induced shower above 1 TeV is expected to be on the order of $50^\circ$~\cite{Resconi:2008fe}. We conservatively consider the signal and background over a solid angle corresponding to a full half of the sky ($2 \pi$ sr), acknowledging that our results would be strengthened if better angular resolution could be obtained. Using an NFW profile, we integrate the dark matter distribution in the direction of the Galactic center over this solid angle. We take the energy resolution of the detector to be $\log({{\rm E_{max}} / {\rm E_{min}}}) \sim 0.3$  \cite{Resconi:2008fe}.

\begin{figure*}[ht]
\includegraphics{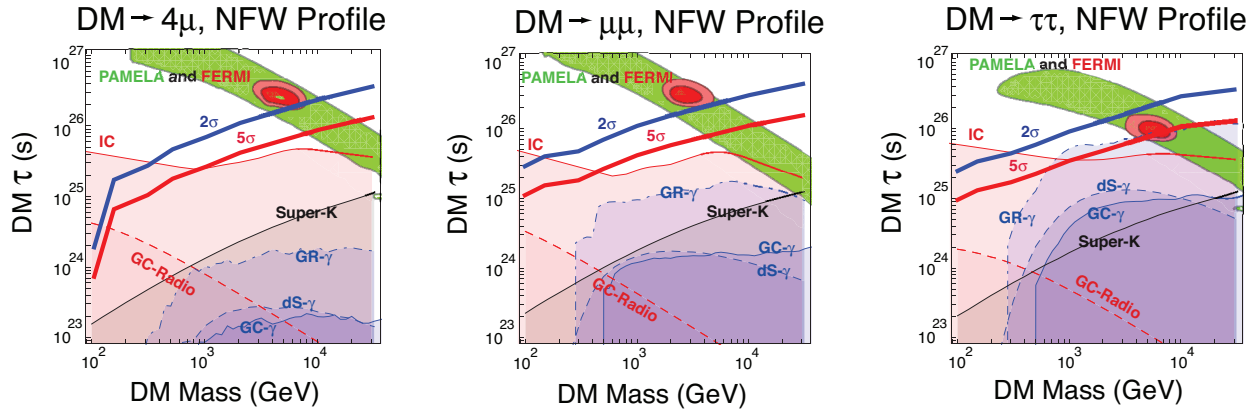}
\caption{Exclusion limits in the dark matter lifetime versus mass plane, for decays to $\mu^+\mu^-$, $\tau^+\tau^-$ and  $\mu^+\mu^-\mu^+\mu^-$. The green region is preferred by the observations of PAMELA, light red is preferred by FGST, and the dark red region is preferred by the combination of these measurements. Here, we assume that the excess of the cosmic $e^+ +e^-$ spectrum measured by FGST beyond the simple power law is the dark matter decay signal (For alternative interpretations, see, {\it e.g.}\/, \cite{Grasso:2009ma}). The blue line labeled ``$2\sigma$'' is the projected $2\sigma$ limit for 5-years of running at IceCube/DeepCore, while the red ``$5\sigma$'' line is the $5\sigma$ discovery reach, assuming an NFW profile (the results change only mildly if another profile is adopted). The black line labeled ``Super-K'' is the limit from neutrino observations at Super-Kamiokande~\cite{Desai:2007ra,Desai:2004pq}. HESS bounds from the Galactic Ridge (``GR-$\gamma$'')~\cite{HessGR}, Galactic Center (``GC-$\gamma$")~\cite{HessGC}, and dwarf spheroidals (``dS-$\gamma$'')~\cite{HessSgrDwarf,VERITAS} are also displayed. The dS-$\gamma$ limit originally calculated in \cite{Essig:2009jx}. Inverse Compton scattering limits from FGST are labeled as ``IC,'' and radio observations of the Galactic Center are ``GC-radio''~\cite{radio}. Our results (blue $2\sigma$ and red 5$\sigma$ lines from IceCube/DeepCore) have been superimposed upon a plot taken from Ref.~\cite{Meade:2009iu} with permission of the authors. \label{fig:preferred}}
\end{figure*}

For dark matter masses between $100$~GeV and $30$~TeV, we calculate the lifetime for which IceCube/DeepCore would provide either a $2\sigma$ exclusion limit or a $5\sigma$ discovery after 5 years of observation. The limits for the three decay channels under consideration are shown in Table~\ref{tab:data}. In Fig.~\ref{fig:preferred}, we overlay our results on top of the regions of $m_\chi$ vs.~$\tau$ parameter space preferred by PAMELA/FGST. Also shown are the regions of parameter space excluded by HESS and VERITAS~\cite{Bertone:2008xr,HessGC,HessGR,HessSgrDwarf,VERITAS}
as well as the limits from Super-Kamiokande~\cite{Desai:2007ra,Desai:2004pq},
FGST observations of gamma rays from inverse Compton scattering below $10~$GeV, and radio observations of the Galactic Center. 
In the figure, our results (blue and red 2 and 5$\sigma$ lines from IceCube/DeepCore) have been superimposed upon a plot taken from Ref.~\cite{Meade:2009iu} with permission of
the authors. In this plot, the dwarf spheroidal (dS-$\gamma$) limit was originally determined in Ref.~\cite{Essig:2009jx}.

A noticeable dip in sensitivity is found for the case of a $\sim$100~GeV dark matter particle decaying to four muons. In this case, the maximum energy carried away by each muon is not much larger than the energy threshold of IceCube/DeepCore, approximately 20 GeV. As a result, most of the neutrinos from such decays are unobservable.

\begin{table}[ht]

\begin{tabular}{|c|c|c|c|}
\hline
$m_\chi$ & Bin Size & $5\sigma$ Detection & $2\sigma$ Exclusion \\
(GeV) & (GeV) & $(\chi \to \mu^+\mu^-)$ & $(\chi \to \mu^+\mu^-)$ \\ \hline \hline
100 & $20-250$ & $\tau < 0.11 \times 10^{26}$~s & $\tau < 0.28 \times 10^{26}$~s \\ \hline
150 & $20-250$ & $\tau < 0.15 \times 10^{26}$~s & $\tau < 0.38 \times 10^{26}$~s \\ \hline
300 & $20-250$ & $\tau < 0.18 \times 10^{26}$~s & $\tau < 0.45 \times 10^{26}$~s \\ \hline
500 & $100-500$ & $\tau < 0.26 \times 10^{26}$~s & $\tau < 0.65 \times 10^{26}$~s \\ \hline
1000 & $150-800$ & $\tau < 0.40 \times 10^{26}$~s & $\tau < 1.0 \times 10^{26}$~s \\ \hline
2000 & $300-1500$ & $\tau < 0.55 \times 10^{26}$~s & $\tau < 1.4 \times 10^{26}$~s \\ \hline
10000 & $1500-8000$ & $\tau < 1.0 \times 10^{26}$~s & $\tau < 2.7 \times 10^{26}$~s \\ \hline
30000 & $5000-25000$ & $\tau < 1.4 \times 10^{26}$~s & $\tau < 3.9 \times 10^{26}$~s \\ \hline
\end{tabular}

\begin{tabular}{|c|c|c|c|}
\hline
$m_\chi$ & Bin Size & $5\sigma$ Detection & $2\sigma$ Exclusion \\
(GeV) & (GeV) & $(\chi \to \tau^+\tau^-)$ & $(\chi \to \tau^+\tau^-)$ \\ \hline \hline
100 & $20-250$ & $\tau < 0.099 \times 10^{26}$~s & $\tau < 0.25 \times 10^{26}$~s \\ \hline
150 & $20-250$ & $\tau < 0.13 \times 10^{26}$~s & $\tau < 0.34 \times 10^{26}$~s \\ \hline
300 & $20-250$ & $\tau < 0.17 \times 10^{26}$~s & $\tau < 0.43 \times 10^{26}$~s \\ \hline
500 & $100-500$ & $\tau < 0.23 \times 10^{26}$~s & $\tau < 0.57 \times 10^{26}$~s \\ \hline
1000 & $150-800$ & $\tau < 0.34 \times 10^{26}$~s & $\tau < 0.87 \times 10^{26}$~s \\ \hline
2000 & $300-1500$ & $\tau < 0.47 \times 10^{26}$~s & $\tau < 1.2\times 10^{26}$~s \\ \hline
10000 & $1500-8000$ & $\tau < 1.0 \times 10^{26}$~s & $\tau < 2.7 \times 10^{26}$~s \\ \hline
30000 & $5000-25000$ & $\tau < 1.2 \times 10^{26}$~s & $\tau < 3.3 \times 10^{26}$~s \\ \hline
\end{tabular}

\begin{tabular}{|c|c|c|c|}
\hline
$m_\chi$ & Bin Size & $5\sigma$ Detection & $2\sigma$ Exclusion \\
(GeV) & (GeV) & $(\chi \to \mu^+\mu^-\mu^+\mu^-)$ & $(\chi \to \mu^+\mu^-\mu^+\mu^-)$ \\ \hline \hline
100 & $20-250$ & $\tau <0.0081 \times 10^{26}$~s & $\tau < 0.021  \times 10^{26}$~s \\ \hline
150 & $20-250$ & $\tau <0.068  \times 10^{26}$~s & $\tau < 0.17 \times 10^{26}$~s \\ \hline
300 & $20-250$ & $\tau <0.11  \times 10^{26}$~s & $\tau < 0.27 \times 10^{26}$~s \\ \hline
500 & $20-250$ & $\tau <0.18  \times 10^{26}$~s & $\tau < 0.44 \times 10^{26}$~s \\ \hline
1000 & $100-500$ & $\tau <0.25  \times 10^{26}$~s & $\tau < 0.65 \times 10^{26}$~s \\ \hline
2000 & $150-800$ & $\tau <0.40 \times 10^{26}$~s & $\tau < 1.0 \times 10^{26}$~s \\ \hline
10000 & $600-3000$ & $\tau <0.82  \times 10^{26}$~s & $\tau < 2.1 \times 10^{26}$~s \\ \hline
30000 & $1500-8000$ & $\tau <1.2  \times 10^{26}$~s & $\tau < 3.2\times 10^{26}$~s \\ \hline
\end{tabular}
\caption{Limits on lifetime, $\tau$, of dark matter decaying into $\mu^+\mu^-$ (top), $\tau^+\tau^-$ (middle), or $\mu^+\mu^-\mu^+\mu^-$ (bottom) after 5 years of data collection at IceCube. Limits are placed for a discovery at $5\sigma$ as well as a bound for $2\sigma$ exclusion.  \label{tab:data}}
\end{table}

From the results of Fig.~\ref{fig:preferred}, we see that IceCube, including the DeepCore extension, can probe the most interesting regions of the $m_\chi$-$\tau$ parameter space with five years of data after the planned completion in 2010. Due to its vastly larger volume, IceCube can probe dark matter decay lifetimes that are orders of magnitude greater than those excluded by Super-Kamiokande. IceCube/DeepCore will be 
competitive with or more constraining than inverse Compton and $\gamma$-ray measurements, and will serve as a complementary test of the dark matter interpretation of PAMELA and FGST results.

KF is supported by the US Department of Energy and MCTP via the Univ.\ of
Michigan and the National Science Foundation under Grant No. PHY-0455649; DS is supported by NSF grant AST-0507117 and GAANN (D.S.); DH is supported by the US Department of Energy, including grant DE-FG02-95ER40896, and by NASA grant NAG5-10842; MRB is supported by the Department of Energy, under grant DE-FG03-92-ER40701. HM is supported in part by World Premier International Research Center Initiative (WPI Initiative), MEXT, Japan, in part by the U.S. DOE under Contract DE-AC03-76SF00098, and in part by the NSF under grant PHY-04-57315. The authors would also like to thank the Aspen Center for Physics for providing a stimulating atmosphere for research and collaboration. DS would also like to thank the MCTP.


\begin{thebibliography}{99}
\bibitem{Adriani:2008zr}
  O.~Adriani {\it et al.}  [PAMELA Collaboration],
  Nature {\bf 458}, 607 (2009)
  [arXiv:0810.4995 [astro-ph]].

\bibitem{Chang:2008zzr}
  J.~Chang {\it et al.},
  Nature {\bf 456}, 362 (2008).
  
\bibitem{Torii:2008xu}
  S.~Torii {\it et al.}  [PPB-BETS Collaboration],
  arXiv:0809.0760 [astro-ph].

\bibitem{Collaboration:2009zk}
  A.~A.~Abdo {\it et al.}  [The Fermi LAT Collaboration],
  Phys.\ Rev.\ Lett.\  {\bf 102}, 181101 (2009)
  [arXiv:0905.0025 [astro-ph.HE]].

\bibitem{Barwick:1997ig}
  S.~W.~Barwick {\it et al.}  [HEAT Collaboration],
  Astrophys.\ J.\  {\bf 482}, L191 (1997)
  [arXiv:astro-ph/9703192].
  
\bibitem{ams}
  M.~Aguilar {\it et al.}  [AMS-01 Collaboration],
  Phys.\ Lett.\  B {\bf 646}, 145 (2007)
  [arXiv:astro-ph/0703154].


 


  
\bibitem{annihilation}
  E.~A.~Baltz, J.~Edsjo, K.~Freese and P.~Gondolo,
  Phys.\ Rev.\  D {\bf 65}, 063511 (2002)
  [arXiv:astro-ph/0109318];
  V.~Barger, W.~Y.~Keung, D.~Marfatia and G.~Shaughnessy,
  arXiv:0809.0162 [hep-ph];



\bibitem{weiner}
  I.~Cholis, L.~Goodenough, D.~Hooper, M.~Simet and N.~Weiner,
  arXiv:0809.1683 [hep-ph];

  \bibitem{Bergstrom:2009fa}
  L.~Bergstrom, J.~Edsjo and G.~Zaharijas,
  arXiv:0905.0333 [astro-ph.HE].

\bibitem{leptonic}
  R.~Harnik and G.~D.~Kribs,
  arXiv:0810.5557 [hep-ph];
  A.~E.~Nelson and C.~Spitzer,
  arXiv:0810.5167 [hep-ph];
  I.~Cholis, D.~P.~Finkbeiner, L.~Goodenough and N.~Weiner,
  arXiv:0810.5344 [astro-ph];
  K.~M.~Zurek,
  arXiv:0811.4429 [hep-ph];
  P.~J.~Fox and E.~Poppitz,
  arXiv:0811.0399 [hep-ph];
  C.~R.~Chen and F.~Takahashi,
  arXiv:0810.4110 [hep-ph];
  I.~Cholis, G.~Dobler, D.~P.~Finkbeiner, L.~Goodenough and N.~Weiner,
  arXiv:0811.3641 [astro-ph].
  D.~Hooper and K.~M.~Zurek,
  arXiv:0902.0593 [hep-ph];
  D.~Hooper and T.~M.~P.~Tait,
  arXiv:0906.0362 [hep-ph].

\bibitem{Sommerfeld1} 
A. Sommerfeld, Annalen der Physik 403, 257 (1931);  
  N.~Arkani-Hamed, D.~P.~Finkbeiner, T.~Slatyer and N.~Weiner,
  arXiv:0810.0713 [hep-ph].

\bibitem{Sommerfeld2}
  M.~Cirelli and A.~Strumia,
  arXiv:0808.3867 [astro-ph];

\bibitem{BW} 
  M.~Ibe, H.~Murayama and T.~T.~Yanagida,
  Phys.\ Rev.\  D {\bf 79}, 095009 (2009)
  [arXiv:0812.0072 [hep-ph]].

\bibitem{decay}
  A.~Arvanitaki, S.~Dimopoulos, S.~Dubovsky, P.~W.~Graham, R.~Harnik and S.~Rajendran,
  arXiv:0812.2075 [hep-ph];
  E.~Nardi, F.~Sannino and A.~Strumia,
  JCAP {\bf 0901}, 043 (2009)
  [arXiv:0811.4153 [hep-ph]];
  A.~Arvanitaki, S.~Dimopoulos, S.~Dubovsky, P.~W.~Graham, R.~Harnik and S.~Rajendran,
  arXiv:0904.2789 [hep-ph].
  
  
  
\bibitem{Profumo:2008ms}
  D.~Hooper, P.~Blasi and P.~D.~Serpico,
  JCAP {\bf 0901}, 025 (2009)
  [arXiv:0810.1527 [astro-ph]];
  S.~Profumo,
  arXiv:0812.4457 [astro-ph].
  



\bibitem{Primack:1988zm}
  J.~R.~Primack, D.~Seckel and B.~Sadoulet,
  Ann.\ Rev.\ Nucl.\ Part.\ Sci.\  {\bf 38}, 751 (1988);
  G.~Jungman, M.~Kamionkowski and K.~Griest,
  Phys.\ Rept.\  {\bf 267}, 195 (1996)
  [arXiv:hep-ph/9506380];
  J.~D.~Lewin and P.~F.~Smith,
  Astropart.\ Phys.\  {\bf 6}, 87 (1996);
  G.~Bertone, D.~Hooper and J.~Silk,
  Phys.\ Rept.\  {\bf 405}, 279 (2005)
  [arXiv:hep-ph/0404175].
  


\bibitem{antiprotons}
  M.~Cirelli, M.~Kadastik, M.~Raidal and A.~Strumia,
  Nucl.\ Phys.\  B {\bf 813}, 1 (2009)
  [arXiv:0809.2409 [hep-ph]];
  F.~Donato, D.~Maurin, P.~Brun, T.~Delahaye and P.~Salati,
  arXiv:0810.5292 [astro-ph].
  
\bibitem{Ishiwata:2009vx}
  K.~Ishiwata, S.~Matsumoto and T.~Moroi,
  JHEP {\bf 0905}, 110 (2009)
  [arXiv:0903.0242 [hep-ph]].

\bibitem{Meade:2009iu}
  P.~Meade, M.~Papucci, A.~Strumia and T.~Volansky,
  arXiv:0905.0480 [hep-ph].
  
\bibitem{Ibarra:2009dr}
  A.~Ibarra, D.~Tran and C.~Weniger,
  arXiv:0906.1571 [hep-ph].
  

\bibitem{Spolyar:2009kx}
  D.~Spolyar, M.~Buckley, K.~Freese, D.~Hooper and H.~Murayama,
  arXiv:0905.4764 [astro-ph.CO].
  
\bibitem{Klein:2008px}
  S.~R.~Klein  [IceCube Collaboration],
  IEEE Trans.\ Nucl.\ Sci.\  {\bf 56}, 1141 (2009)
  [arXiv:0807.0034 [physics.ins-det]].
  
\bibitem{Resconi:2008fe}
  E.~Resconi [IceCube Collaboration],
  Nucl.\ Instrum.\ Meth.\  A {\bf 602}, 7 (2009)
  [arXiv:0807.3891 [astro-ph]].
  
\bibitem{Jungman:1994jr}
G.~Jungman and M.~Kamionkowski,
Phys.\ Rev.\ D {\bf 51} (1995) 328
[arXiv:hep-ph/9407351];
  D.~Grellscheid and P.~Richardson,
  arXiv:0710.1951 [hep-ph].
  
\bibitem{Sjostrand:2006za}
  T.~Sjostrand, S.~Mrenna and P.~Skands,
  JHEP {\bf 0605}, 026 (2006)
  [arXiv:hep-ph/0603175].
  


  
\bibitem{Honda:2006qj}
  M.~Honda, T.~Kajita, K.~Kasahara, S.~Midorikawa and T.~Sanuki,
  Phys.\ Rev.\  D {\bf 75}, 043006 (2007)
  [arXiv:astro-ph/0611418];
  J.~G.~Learned and K.~Mannheim,
  Ann.\ Rev.\ Nucl.\ Part.\ Sci.\  {\bf 50}, 679 (2000);
see also, G. Barr {\it et al}, Phys. Rev. D. {\bf 70}, 023006 (2004).
  


\bibitem{Collaboration:2009nf}
  The IceCube Collaboration,
  arXiv:0902.0675 [astro-ph.HE].
  
\bibitem{Gandhi:1998ri}
  R.~Gandhi, C.~Quigg, M.~H.~Reno and I.~Sarcevic,
  Phys.\ Rev.\  D {\bf 58}, 093009 (1998)
  [arXiv:hep-ph/9807264].
    
\bibitem{Bertone:2008xr}
  G.~Bertone, M.~Cirelli, A.~Strumia and M.~Taoso,
  JCAP {\bf 0903}, 009 (2009)
  [arXiv:0811.3744 [astro-ph]].



  
  
  \bibitem{HessGC}
Matthieu Vivier, talk presented on behalf of 
HESS collaboration, 44th Rencontres de Moriond,
February 6, 2009.
%



\bibitem{HessGR}  
HESS collaboration,
Nature {439} (2006) 695
[astro-ph/0603021].


\bibitem{HessSgrDwarf}
HESS collaboration, Astropart. Phys. {29} (2008) 55 [arXiv:0711.2369]. 


\bibitem{VERITAS}
C.~M.~Hui and f.~t.~V.~Collaboration,
AIP Conf.\ Proc.\  {\bf 1085} (2009) 407
[arXiv:0810.1913 [astro-ph]].

\bibitem{Essig:2009jx}
  R.~Essig, N.~Sehgal and L.~E.~Strigari,
  Phys.\ Rev.\  D {\bf 80}, 023506 (2009)
  [arXiv:0902.4750 [hep-ph]].

\bibitem{Grasso:2009ma}
  D.~Grasso {\it et al.}  [FERMI-LAT Collaboration],
  arXiv:0905.0636 [astro-ph.HE].

   
  \bibitem{Desai:2007ra}
  S.~Desai {\it et al.}  [Super-Kamiokande Collaboration],
  Astropart.\ Phys.\  {\bf 29}, 42 (2008)
  [arXiv:0711.0053 [hep-ex]].
  
  \bibitem{Desai:2004pq}
  S.~Desai {\it et al.}  [Super-Kamiokande Collaboration],
  Phys.\ Rev.\  D {\bf 70}, 083523 (2004)
  [Erratum-ibid.\  D {\bf 70}, 109901 (2004)]
  [arXiv:hep-ex/0404025].


\bibitem{radio}
R.D.Davis, D.Walsh, R.S.Booth, MNRAS 177, 319-333 (1976)
R.~Genzel {\it et al.},
 Nature {425} (2003) 934
 [astro-ph/0310821].

 
  


\end{thebibliography}
\end{document}